\documentclass[prl,twocolumn,english,aps,reprint,superscriptaddress,notitlepage,nobibnote,preprintnumbers,showpacs]{revtex4-1}
\usepackage{graphicx}
\usepackage{dcolumn}
\usepackage{bm}
\usepackage{epsfig}
\usepackage{gensymb}
\usepackage{mathrsfs}
\usepackage{amsmath}
\usepackage{amssymb}
\usepackage{graphicx,bm}
\usepackage{natbib}
\usepackage{slashed}
\usepackage[makeroom]{cancel}
\usepackage{dsfont}
\usepackage{longtable}
\usepackage{multirow}

 \usepackage[titletoc]{appendix}
 
\usepackage{titlesec}
\titleformat{\section}[runin]{\normalfont \bfseries}{\thesection}{1 em}{}
\titleformat{\subsection}[runin]{\normalfont \bfseries}{\thesubsection}{1em}{}

\titlespacing{\subsection}{0pt}{5.0pt}{15.5pt}

\titlespacing{\section}{0pt}{5.0pt}{15.5pt}

\def\fun#1#2{\lower3.6pt\vbox{\baselineskip0pt\lineskip.9pt
  \ialign{$\mathsurround=0pt#1\hfil##\hfil$\crcr#2\crcr\sim\crcr}}}

\usepackage{tikz}
\usepackage{environ}
\makeatletter
\newsavebox{\measure@tikzpicture}
\NewEnviron{scaletikzpicturetowidth}[1]{%
  \def\tikz@width{#1}%
  \begin{lrbox}{\measure@tikzpicture}%
  \BODY
  \end{lrbox}%
  \pgfmathparse{#1/\wd\measure@tikzpicture}%
  \BODY
}
\makeatother
\usetikzlibrary{decorations.pathmorphing,decorations.markings,trees,shapes}

\def\lsim{\mathrel{\rlap{\raise 2.5pt \hbox{$<$}}\lower 2.5pt\hbox{$\sim$}}}
\def\gsim{\mathrel{\rlap{\raise 2.5pt \hbox{$>$}}\lower 2.5pt\hbox{$\sim$}}}

\newcommand{\bea}{\begin{eqnarray}}
\newcommand{\eea}{\end{eqnarray}}

\input epsf

\newcommand{\blue}[1]{\textcolor{blue}{#1}}
\usepackage{color}

\newcommand{\comment}[1]{}

\newcommand{\mc}[1]{\mathcal{#1}}

\newcommand{\vev}[1]{\langle #1 \rangle}
\newcommand{\ket}[1]{| #1 \rangle}

\newcommand{\inner}[2]{\langle #1 | #2 \rangle}

\newcommand{\eq}[1]{\begin{equation}\begin{split} #1 \end{split}\end{equation}}
\newcommand{\eqs}[1]{\begin{align} #1 \end{align}}

\begin{document}

\title{Constructing the general partial waves and renormalization in EFT}

\author{Jing Shu}
\email{jshu@pku.edu.cn}
\affiliation{School of Physics and State Key Laboratory of Nuclear Physics and Technology, Peking University, Beijing 100871, China}
\affiliation{Center for High Energy Physics, Peking University, Beijing 100871, China}
\affiliation{School of Fundamental Physics and Mathematical Sciences, Hangzhou Institute for Advanced\\ Study, University of Chinese Academy of Sciences, Hangzhou 310024, China}
\affiliation{International Center for Theoretical Physics Asia-Pacific, Beijing/Hangzhou, China}
\author{Ming-Lei Xiao}
\email{minglei.xiao@northwestern.edu}
\affiliation{Northwestern University, Evanston, Illinois 60208, USA.}
\affiliation{Argonne National Laboratory, Lemont, Illinois 60439, USA.}
\author{Yu-Hui Zheng}
\email{zhengyuhui@itp.ac.cn}
\affiliation{
CAS Key Laboratory of Theoretical Physics, Institute of Theoretical Physics,  Chinese Academy of Sciences, Beijing 100190, P. R. China.}
\affiliation{School of Physical Sciences, University of Chinese Academy of Sciences, Beijing 100049, China.}

\begin{abstract}\indent
We construct the general partial wave amplitude basis for the $N\to M$ scattering, which consists of Poincar\'e Clebsch-Gordan coefficients, with Lorentz invariant forms given in terms of spinor-helicity variables. The inner product of the Clebsch-Gordan coefficients is defined, which converts on-shell phase space integration into an algebraic problem.
We also develop the technique of partial wave expansions of arbitrary amplitudes, including those with infrared divergence.
These are applied to the computation of anomalous dimension matrix for general effective operators, where unitarity cuts for the loop amplitudes, with an arbitrary number of external particles, are obtained via partial wave expansion.

\end{abstract}

\maketitle

\section{Introduction}

Partial wave analysis has been an important technique since the era of classical field theory, and remains important in modern scattering theories. 
Recently, there has been raising interest in theoretical constraints on effective field theories from the principle of unitarity, while partial wave analysis tends to provide a tighter bound than a general analysis \cite{Arkani-Hamed:2020blm,Wang:2020xlt}. 
There are also observations of new selection rules and computational techniques \cite{Jiang:2020rwz,Baratella:2020dvw} for anomalous dimensions based on partial waves.
It is hence instructive to implement a complete study on the partial waves in general for scattering problems.

Traditional partial waves are defined as eigenfunctions of the angular momentum operator ${\bf J}^2$, particularly for two-particle systems in the Center of Mass (CoM) frame. 
From the covariant point of view \cite{Jiang:2020rwz}, the partial wave functions form irreducible representations of the Poincar\'e group, which are eigenfunctions of the Casimir operators ${\bf P}^2$ and ${\bf W}^2$, where ${\bf W}_{\mu} = \frac12\epsilon_{\mu\nu\rho\sigma}{\bf P}^{\nu}{\bf M}^{\rho\sigma}$ is the Pauli-Lubanski operator. 
In the CoM frame $P^{\mu} = (E,0,0,0)$, ${\bf W}^2$ reduces to $E^2{\bf J}^2$, so that the eigenfunctions reduce to the traditional partial wave functions as expected. 
To construct the covariant partial wave functions, one could start from the Poincar\'e irreducible multi-particle state $\ket{P,J,\sigma}$ as the eigenstate of the Casimir, where $\sigma$ is the helicity. The Poincar\'e Clebsch-Gordan Coefficients (CGC) of a multi-particle state $\ket{\{\Psi\}^N}$ is defined as the overlap
\eq{
    \inner{P,J,\sigma}{\Psi_1,\dots,\Psi_N} = \mc{C}^{J,\sigma}_{N} \delta^{(4)}(P-\sum_{i=1}^N p_i)\,,
}
which constitute the partial wave functions as
\eq{\label{eq:pw_cgc}
	\mc{B}^J_{i\to f} = \sum_{\sigma} \mathcal{C}_i^{J,\sigma} (\mathcal{C}_f^{J,\sigma})^* \equiv \mathcal{C}_i^{J} \cdot \mathcal{C}_{\bar{f}}^{J} \,.
}
where $\bar{f}$ means the CP conjugate of state $f$.
The explicit form of the CGC can be inferred from decay amplitudes of an auxiliary massive particle \cite{Jiang:2020rwz}, which can be obtained by using the spinor helicity variables \cite{Arkani-Hamed:2017jhn}. For 2-particle states, it recovers the Wigner-D matrix, but the above definition could be generalized to arbitrary scattering.

In this paper, we carry out a closer analysis of the Poincar\'e CGC. First, we normalize them by defining an inner product $\langle\mathcal{C},\mathcal{C}\rangle$, which is
basically an integration over the multi-particle phase space. Then we get an algebraic formula for the phase space integration.
The orthogonality indicated by the factor $\delta^{JJ'}$ is precisely the selection rules in \cite{Jiang:2020rwz}.
In order to apply it to phase space integrations of physical amplitudes $\mc{A}$, we need to perform the partial wave expansion $\mc{A} = \sum_J \mc{M}^J(s) \mc{B}^J$.
There are three classes of amplitudes: i) for local amplitudes, we use the Young Tableau basis to find the coordinates of both the partial wave basis and the amplitude to be decomposed, so that the linear expansion can be solved; ii) for non-local amplitudes without IR divergence, we explicitly perform the phase space integration in terms of the spinor variables; iii) there are also amplitudes with diverging phase space integration, such as the Coulomb scattering, for which we need to use the BCFW shift to eliminate the IR divergence. 
This technique can be directly applied to compute the unitarity cut of one-loop amplitudes,
which gives the renormalization of the corresponding j-basis operator $\mc{O}^{J}$. 
Therefore, as long as the partial wave coefficients $\mc{M}^J$ are worked out for various tree-level amplitudes, we can provide an algebraic method of computing the Anomalous Dimension Matrix (ADM) for the effective operators. 
%


\section{Phase Space Integration for Partial Waves}\label{sec:GPWB}

In \cite{Jiang:2020rwz}, the partial wave functions are defined via the Poincar\'e CGC as in eq.~\eqref{eq:pw_cgc}, without a specific normalization. On the other hand, we are familiar with how traditional partial wave functions are normalized by a spatial integration, such as
\eq{\label{eq:wignerD}\small
    \int d\Omega\, D^{J}_{\Delta\sigma}(\Omega)D^{J'}_{\Delta\sigma'}(\Omega)^* =\frac{4\pi}{2J+1}\delta^{JJ'}\delta^{\sigma\sigma'}\,.
}
Thus it is natural to define a normalization of the CGC by a phase space integration, which stems from the orthonormality of multi-particle states
\begin{align}\small
    &\langle P,J,\sigma|P',J',\sigma'\rangle\equiv \;\Big\langle \mathcal{C}^{J,\sigma}_N , \;\mathcal{C}^{J',\sigma'}_N \Big\rangle \delta^4(P-P') \notag\\
    &=\int d\Phi_N \, \langle P,J,\sigma|\{\Psi_i\}^N\rangle\langle \{\Psi_i\}^N|P',J',\sigma'\rangle,
\end{align}
where $d\Phi_N$ is the measure of $N$-body phase space, and we have defined an inner product among the CGC of the same $N$-particles as~\footnote{
When $N>2$, there are degenerate CGC's for a fixed angular momentum $J$, which we may label by an additional index as $\mc{C}^{J,\sigma,a}_N$. In this case, the inner product yields a matrix in the degenerate subspace, denoted as $g^{aa'}_N(J)$, which is the metric defined by the inner product.
}
\begin{align}\label{eq:cgc_inner}
\small\begin{split}
    \Big\langle \mathcal{C}^{J,\sigma}_N , \;\mathcal{C}^{J',\sigma'}_N \Big\rangle
    &=\int d\Phi_N \, \mathcal{C}^{J,\sigma}\,(\mathcal{C}^{J',\sigma'})^* = g_N(J)\delta^{JJ'}\delta^{\sigma\sigma'}
\end{split}\end{align}
where the Kronecker deltas are predicted by the angular momentum conservation. 
When $N>2$, there are degenerate CGC's for a fixed angular momentum $J$, which we may label by an additional index as $\mc{C}^{J,\sigma,a}$. In this case, the inner product yields a matrix in the degenerate subspace, denoted as $g^{aa'}_N(J)$, which is a metric defined by the inner product.

To explicitly compute the inner product, we make use of the spinor helicity variables. The CGC has the form of a decay amplitude, which can be written in terms of the spinor variables, such as for 2-particle states
\eq{\label{eq:cgc_2}
    \mc{C}^{J,\sigma}_{h_1,h_2} &= \frac{[12]^{h_1+h_2-J}}{\sqrt{\mathcal{N}}s^{(J+h_1+h_2)/2}} \\
    &\times ([1\bm{x}]^{J+h_1-h_2}[2\bm{x}]^{J-h_1+h_2})^{\{I_1\dots I_{2J}\}},
}
where $\bm{x}^I$ is the spinor variable for the total momentum $P$, and the little group (LG) indices $I_i$ are totally symmetrized, representing $2J+1$ values of $\sigma$. Herein, we have chosen a particular normalization, where a power of $s=P^2$ ensures that the CGCs remain dimensionless and $\mathcal{N}$ is the number of terms due to the total symmetry of little group indices, such that in the CoM frame, the value of the CGC is given by the Wigner D function
\eq{
    \mc{C}^{J,\sigma}_{h_1,h_2} = D^{J}_{h_1-h_2,\sigma}(\theta,\phi).
}
Given the relation $\int d\Phi_2 = \frac18\int d\Omega$, eq.~\eqref{eq:wignerD} directly induces the inner product in eq.~\eqref{eq:cgc_inner} and yields
\eq{\label{eq:metric}
    g_2(J) = \frac{\pi}{2(2J+1)}\,.
}

To generalize the idea to scattering amplitudes, performing the phase space integration in the CoM frame is not always convenient, especially when there are more than two particles. Hence, we shall use the integral measure of the Lorentz invariant phase space $d\Phi_N$ in terms of spinor variables. For two particles $d\Phi_2$ is given in \cite{Mastrolia:2009dr} as follows: Suppose the momenta of the two particles are $k'_1$, $k'_2$, with $P=k'_1+k'_2$. Define $(k'_1)_{\alpha\dot\alpha}=t|l\rangle_{\alpha}[l|_{\dot{\alpha}}$, so that the integral can be expressed as
\begin{align}
    \begin{split}
    \int d\Phi_{1',2'}=&\int d^4k'_1\,\delta\left((k'_1)^2\right)\,\delta\left((P-k'_1)^2\right)\\
    =\frac{i}{4}\int &tdt\,\langle l\, dl\rangle[dl\, l]\,\delta\left( 2t\, \vev{l\bm{x}}[l\bm{x}]+P^2 \right).
    \end{split}
\end{align}
It can be converted to a complex integral by expanding the loop spinors in a fixed basis $|l\rangle=|1\rangle+z|2\rangle$, $|l]=|1]+\bar{z}|2]$ with $P=\ket{1}[1|+\ket{2}[2|$, as
\begin{align}\begin{split}\label{eq:int_2}
    \int d\Phi_{1',2'}
    =\frac{i}{4}\oint dz\int d\bar{z}\, dt\, t^2\,\delta\left(t+\frac{1}{1+z\bar{z}}\right).
\end{split}\end{align}
Apply this integration to eq.~\eqref{eq:cgc_inner} with $N=2$ and plug in eq.~\eqref{eq:cgc_2}, we get
\eq{
    \int d\Phi_{1',2'}\,\mc{C}^{J,\sigma}_{h'_1,h'_2}(\mc{C}^{J',\sigma'}_{h'_1,h'_2})^* = \frac{\pi \delta^{JJ'}}{2(2J+1)}\underbrace{\frac{(\delta_{\ I'}^{I})^{2J}_{\rm sym}}{\mc{N}}}_{\simeq\delta^{\sigma\sigma'}}\,,
}
which verifies the desired result in eq.~\eqref{eq:metric}.
The angular momentum conservation is explicitly verified, as when $J\neq J'$ the integration should have factors of $\epsilon^{I_iI_j}$ that contradict the total symmetry of the LG indices.
As a result, we can easily derive the following phase space integration for partial-wave basis defined as eq.~\eqref{eq:pw_cgc}
\eqs{\label{eq:PWBIntegral}
    \int d\Phi_{1',2'}\;\mathcal{B}^J_{i \to 1',2'}\,\mathcal{B}^{J'}_{1',2'\to f}
    &=\mathcal{C}^J_{i}\cdot \Big\langle \mathcal{C}^{J'}_{1',2'},\; \mathcal{C}^{J}_{1',2'} \Big\rangle \cdot \mathcal{C}^{J'}_{\bar{f}}  \nonumber \\
    &=g_2(J)\delta^{JJ'}\, \mathcal{B}^J_{i\to f} \,.
}
which applies to arbitrary initial and final states $\ket{i},\ket{f}$, including three or more particle states. 
For the phase space of more than two particles, the measure can be defined recursively as
\begin{align}
    \int d\Phi_{k'_1,k'_2,k'_3}=\int d(q^2)\; d\Phi_{k'_1,q}\;\delta^{(4)}(q-k'_2-k'_3)\int d\Phi_{k'_2,k'_3}\,,
\end{align}
so that we can normalize the general $N$-particle CGC. This paper only focuses on the two-particle calculations because they are relevant to the one-loop unitarity cut.


\section{Partial-wave expansion }\label{sec:PWE}
With the normalized partial waves, we are ready to do the partial-wave expansion for a general amplitude
\eq{
    \mc{A} = \sum_{J,a}\mc{M}^{J,a}\mc{B}^{J,a}\,,
}
where $\mc{M}^J$ are the partial wave coefficients that may depend on any Mandelstam variables $s_{\mc{I}} = (\sum_{i\in\mc{I}}p_i)^2$ where $\mc{I}$ is a subset of either the initial or final state. The superscript $a$ is for possible degeneracy when the amplitude involves $N>2$ particles in either the initial or final state.

For local amplitudes, we have discussed in \cite{Jiang:2020rwz} that the partial-wave expansion is finite. 
The expansion can be done purely algebraically. With \textbf{ABC4EFT} Mathematica package \cite{Li:2020zfq,Li:2020xlh,Li:2020tsi,Li:2022}, we are allowed to reduce any local amplitudes to the y-basis $\{\mathcal{B}^{\rm y}_i\}$
\begin{align}\label{eq:pw_expand_local}\small
    \begin{array}{l}
    \mathcal{A}=\sum_ic_i\mathcal{B}^{\rm y}_i \\
    \mathcal{B}^{J,a}=\sum_i\mathcal{K}^{\rm (jy)}_{Ja,i}\mathcal{B}^{\rm y}_i
    \end{array}\Bigg\}   \Rightarrow 
    \mc{M}^{J,a}=\sum_{i}c_i(\mathcal{K}^{(\rm jy)})^{-1}_{i,Ja}\,.
\end{align}
where $\mc{K}^{\rm (jy)}$ denotes the transfer matrix between the j-basis (partial wave) and the y-basis. One may refer to \cite{Li:2020zfq,Li:2022abx} for more details.

For non-local amplitudes, such as tree-level amplitudes with poles, we can use eq.~\eqref{eq:PWBIntegral} to extract the partial wave coefficient of $2\to n$ scattering from the following expansion
\eq{\label{eq:pw_expand}
    \int d\Phi_{1',2'} \mc{C}^J_{\bar{1}',\bar{2}'}\mc{A}_{1',2'\to f} = g_2(J)\sum_a\mc{M}^{J,a}\mc{C}^{J,a}_{\bar{f}} \,.
}
The integral is performed with the technique of eq.~\eqref{eq:int_2}, which is, however pathological in some cases due to the IR divergence. The Coulomb scattering is a famous example of the IR divergence, for which partial wave expansion is not viable without proper regularization \cite{Taylor:1973da}. The authors of \cite{Baratella:2020dvw} defined the regularized partial wave coefficient by removing a known piece of IR divergence in the integral for $2\to2$ scattering. It is not enough for arbitrary scattering. Thus we propose to adopt the BCFW trick to remove the most general one-loop IR divergence. In particular, we apply the BCFW shift to the loop momenta $\hat{k}'_1= k_1'+z q$, $\hat{k}'_2=k'_2-zq$ (for example $| \hat{1}'\rangle= |1'\rangle +z|2'\rangle$, $| \hat{2}']= |2'] -z|1']$) to obtain the shifted CGC $\hat{\mc{C}}$ and amplitude $\hat{\mc{A}}$, so that the residue at complex infinity leads to the regularized partial wave
\eq{\label{eq:pw_expand_reg}
    \int d\Phi_{1',2'} \oint_{\infty}\frac{dz}{z}\,\hat{\mc{C}}^J(z)\hat{\mc{A}}(z) = g_2(J)\sum_a\mc{M}^{J,a}_{\rm reg}\mc{C}^{J,a}_{\bar{f}} \,.
}
We provide detailed examples of partial wave expansions for all the cases mentioned above in the supplementary material.

\section{One-loop Anomalous Dimension via On-shell Method}\label{sec:AnomalousDim}
The partial wave expansion can be applied to the computations of one-loop anomalous dimensions of effective operators, which is shown for $2\to2$ scattering in \cite{Baratella:2020dvw}. We generalize the idea to arbitrary scattering so that any operators can be analyzed in this way. 
The UV divergence of the one-loop amplitude only comes from the scalar bubble integrals $I_2=\epsilon^{-1}+O(\epsilon^0)$
\begin{align}\label{eq:PVredu}
    \mathcal{A}^{\rm 1-loop}=\sum_{\mc{I}} b_\mc{I}I_2(s_\mc{I}) + \text{UV finite terms}
\end{align}
where $\mc{I}$ denotes the channel of the bubble, and $s_\mc{I}$ is the Mandelstam variable.
The coefficient $b_\mc{I}$ is given by the double cut unitarity \cite{2004,2005} as
\eq{\label{eq:C2}\small
    & b_\mc{I}=\frac{\text{cut}\mc{A}^{\rm 1-loop}}{\text{cut}I_2(s_\mc{I})} = \frac{2}{\pi} \sum_{1',2'} \int d\Phi_{1',2'} \, \mathcal{A}_{\rm L} \mathcal{A}_{\rm R} \,,
}
where we have summed over all possible intermediate two-particle states $(1',2')$. The sub-amplitudes can be expanded into partial waves
\eq{
    \mc{A}_{\rm L} = \sum_J\mc{M}^J_{\rm L}\mc{B}^J_{i\to 1',2'} \ ,\ 
    \mc{A}_{\rm R} = \sum_J\mc{M}^J_{\rm R}\mc{B}^J_{1',2' \to f}
}
With eq.~\eqref{eq:PWBIntegral} we can directly obtain
\eq{\label{eq:cut_calc}
    b_\mc{I} = \sum_{1',2'}\sum_{J,a,b} \frac{\mc{M}_{\rm L}^{J,a}\mc{M}_{\rm R}^{J,b}}{2J+1}\mc{B}^{J,(a,b)}_\mc{I} \,.
}
Note that the final partial wave is in general labeled by degeneracy from both initial and final states $(a,b)$, which correspond to various independent Lorentz structures in the UV divergence, and hence contribute to the renormalization of independent operators $\mc{O}^{J,(a,b)}$. Such operators with definite angular momentum $J$ in a particular channel are defined as the j-basis operators \cite{Li:2020zfq,Li:2022abx}. 

Now we consider the contribution of the effective operator $\mc{O}_i$ to the renormalization of $\mc{O}_j$, which is characterized by the anomalous dimension matrix $\gamma_{ij}$ defined as
\eq{
    \frac{d C_j}{d \ln\mu} = \sum_i \gamma_{ij}C_i \,.
}
where $C_i$ is the Wilson coefficient of $\mc{O}_i$. With a particular operator basis $\{\mc{O}_i\}$ (f-basis), the expansion of the j-basis operators can be deduced similar to eq.~\eqref{eq:pw_expand_local} as
\eq{
    \mc{O}^{J,a} = \sum_i\mc{K}^{\rm (jf)}_{Ja,i}\mc{O}_i \,.
}
Thus the anomalous dimension can be directly extracted from the above result~\footnote{Regarding the renormalization of an operator by itself, one needs to compute extra contributions from the field renormalization $Z_i$ of the external particles\cite{Jiang:2020mhe}.}
\eq{\label{eq:adm_master}
    \gamma_{ij}C_i = -\frac{1}{8\pi^2}\sum_{\mc{I},1',2'}\sum_{J,a,b}\frac{\mc{M}_{\rm L}^{J,a}\mc{M}_{\rm R}^{J,b}}{2J+1}\mc{K}^{\rm (jf)}_{J(a,b),j} \,.
}
This is the master formula of this paper, which computes the anomalous dimension matrix algebraically if the partial wave expansions are given.
The selection rule \cite{Jiang:2020rwz} is manifest here: when $\mc{M}_{\rm L/R}$ does not have a shared $J$ value with the matrix $\mc{K}^{\rm (jf)}$ for which both are non-zero, we must have $\gamma_{ij} = 0$. It is often the case when $J$ in the subamplitude is fixed by $\mc{O}_i$.

\begin{widetext}

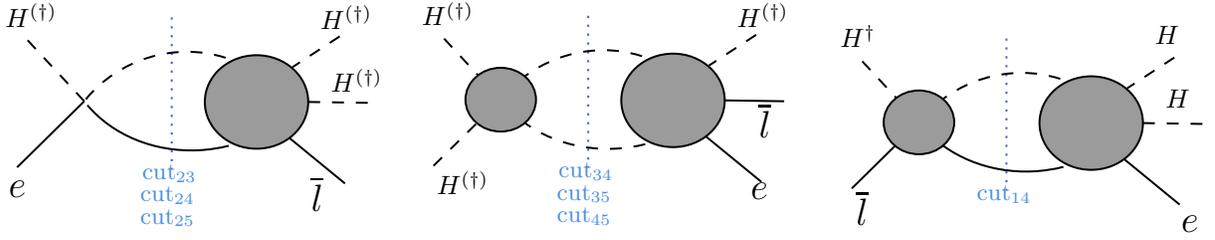
\begin{figure}
    \centering

\tikzset{every picture/.style={line width=0.75pt}} 

\begin{tikzpicture}[x=0.75pt,y=0.75pt,yscale=-1,xscale=1]

\draw    (60.67,90.9) -- (94.56,57.53) ;
\draw  [dash pattern={on 4.5pt off 4.5pt}]  (94.56,57.53) -- (63.31,23.84) ;
\draw [color={rgb, 255:red, 74; green, 88; blue, 226 }  ,draw opacity=1 ] [dash pattern={on 0.84pt off 2.51pt}]  (138.68,15.21) -- (138.68,91.86) ;
\draw  [draw opacity=0] (167.41,79.99) .. controls (162.67,81.2) and (157.7,81.98) .. (152.56,82.24) .. controls (129.51,83.45) and (108.62,74.3) .. (96.16,59.52) -- (149.03,27.29) -- cycle ; \draw   (167.41,79.99) .. controls (162.67,81.2) and (157.7,81.98) .. (152.56,82.24) .. controls (129.51,83.45) and (108.62,74.3) .. (96.16,59.52) ;  
\draw  [draw opacity=0][dash pattern={on 4.5pt off 4.5pt}] (95.35,56.55) .. controls (111.73,36.58) and (140.02,27.34) .. (165.68,35.54) .. controls (167.83,36.22) and (169.91,37.01) .. (171.91,37.9) -- (143.34,90.63) -- cycle ; \draw  [dash pattern={on 4.5pt off 4.5pt}] (95.35,56.55) .. controls (111.73,36.58) and (140.02,27.34) .. (165.68,35.54) .. controls (167.83,36.22) and (169.91,37.01) .. (171.91,37.9) ;  
\draw  [fill={rgb, 255:red, 155; green, 155; blue, 155 }  ,fill opacity=1 ] (155.36,57.89) .. controls (155.36,44.86) and (167.04,34.3) .. (181.44,34.3) .. controls (195.84,34.3) and (207.52,44.86) .. (207.52,57.89) .. controls (207.52,70.92) and (195.84,81.48) .. (181.44,81.48) .. controls (167.04,81.48) and (155.36,70.92) .. (155.36,57.89) -- cycle ;
\draw    (226.05,99.04) -- (197.81,75.89) ;
\draw  [dash pattern={on 4.5pt off 4.5pt}]  (237.53,58.33) -- (207.52,57.89) ;
\draw  [dash pattern={on 4.5pt off 4.5pt}]  (223.41,24.79) -- (199.58,40.32) ;
\draw  [dash pattern={on 4.5pt off 4.5pt}]  (270.71,90.1) -- (304.6,56.73) ;
\draw  [dash pattern={on 4.5pt off 4.5pt}]  (304.6,56.73) -- (273.36,23.04) ;
\draw [color={rgb, 255:red, 74; green, 88; blue, 226 }  ,draw opacity=1 ] [dash pattern={on 0.84pt off 2.51pt}]  (348.72,14.42) -- (348.72,91.06) ;
\draw  [draw opacity=0][dash pattern={on 4.5pt off 4.5pt}] (377.46,79.19) .. controls (372.72,80.41) and (367.74,81.18) .. (362.6,81.45) .. controls (339.55,82.65) and (318.67,73.5) .. (306.2,58.72) -- (359.07,26.49) -- cycle ; \draw  [dash pattern={on 4.5pt off 4.5pt}] (377.46,79.19) .. controls (372.72,80.41) and (367.74,81.18) .. (362.6,81.45) .. controls (339.55,82.65) and (318.67,73.5) .. (306.2,58.72) ;  
\draw  [draw opacity=0][dash pattern={on 4.5pt off 4.5pt}] (305.39,55.75) .. controls (321.77,35.78) and (350.06,26.54) .. (375.72,34.74) .. controls (377.87,35.42) and (379.95,36.22) .. (381.95,37.1) -- (353.39,89.83) -- cycle ; \draw  [dash pattern={on 4.5pt off 4.5pt}] (305.39,55.75) .. controls (321.77,35.78) and (350.06,26.54) .. (375.72,34.74) .. controls (377.87,35.42) and (379.95,36.22) .. (381.95,37.1) ;  
\draw  [fill={rgb, 255:red, 155; green, 155; blue, 155 }  ,fill opacity=1 ] (365.4,57.09) .. controls (365.4,44.06) and (377.08,33.5) .. (391.48,33.5) .. controls (405.89,33.5) and (417.56,44.06) .. (417.56,57.09) .. controls (417.56,70.12) and (405.89,80.68) .. (391.48,80.68) .. controls (377.08,80.68) and (365.4,70.12) .. (365.4,57.09) -- cycle ;
\draw    (436.1,98.25) -- (407.85,75.09) ;
\draw    (447.57,57.53) -- (417.56,57.09) ;
\draw  [dash pattern={on 4.5pt off 4.5pt}]  (433.45,24) -- (409.62,39.52) ;
\draw    (481.63,101.56) -- (515.52,68.19) ;
\draw  [dash pattern={on 4.5pt off 4.5pt}]  (515.52,68.19) -- (484.28,34.49) ;
\draw [color={rgb, 255:red, 74; green, 88; blue, 226 }  ,draw opacity=1 ] [dash pattern={on 0.84pt off 2.51pt}]  (559.65,25.87) -- (559.65,102.52) ;
\draw  [draw opacity=0] (588.38,90.65) .. controls (583.64,91.86) and (578.67,92.63) .. (573.52,92.9) .. controls (550.48,94.11) and (529.59,84.96) .. (517.13,70.17) -- (570,37.95) -- cycle ; \draw   (588.38,90.65) .. controls (583.64,91.86) and (578.67,92.63) .. (573.52,92.9) .. controls (550.48,94.11) and (529.59,84.96) .. (517.13,70.17) ;  
\draw  [draw opacity=0][dash pattern={on 4.5pt off 4.5pt}] (516.32,67.2) .. controls (532.7,47.24) and (560.99,38) .. (586.65,46.19) .. controls (588.8,46.88) and (590.88,47.67) .. (592.88,48.56) -- (564.31,101.29) -- cycle ; \draw  [dash pattern={on 4.5pt off 4.5pt}] (516.32,67.2) .. controls (532.7,47.24) and (560.99,38) .. (586.65,46.19) .. controls (588.8,46.88) and (590.88,47.67) .. (592.88,48.56) ;  
\draw  [fill={rgb, 255:red, 155; green, 155; blue, 155 }  ,fill opacity=1 ] (576.33,68.54) .. controls (576.33,55.51) and (588.01,44.95) .. (602.41,44.95) .. controls (616.81,44.95) and (628.49,55.51) .. (628.49,68.54) .. controls (628.49,81.57) and (616.81,92.14) .. (602.41,92.14) .. controls (588.01,92.14) and (576.33,81.57) .. (576.33,68.54) -- cycle ;
\draw    (647.02,109.7) -- (618.78,86.55) ;
\draw  [dash pattern={on 4.5pt off 4.5pt}]  (658.49,68.98) -- (628.49,68.54) ;
\draw  [dash pattern={on 4.5pt off 4.5pt}]  (644.37,35.45) -- (620.55,50.98) ;
\draw  [fill={rgb, 255:red, 155; green, 155; blue, 155 }  ,fill opacity=1 ] (497.74,68.19) .. controls (497.74,59.3) and (505.7,52.1) .. (515.52,52.1) .. controls (525.34,52.1) and (533.31,59.3) .. (533.31,68.19) .. controls (533.31,77.07) and (525.34,84.27) .. (515.52,84.27) .. controls (505.7,84.27) and (497.74,77.07) .. (497.74,68.19) -- cycle ;
\draw  [fill={rgb, 255:red, 155; green, 155; blue, 155 }  ,fill opacity=1 ] (286.82,56.73) .. controls (286.82,47.85) and (294.78,40.64) .. (304.6,40.64) .. controls (314.42,40.64) and (322.38,47.85) .. (322.38,56.73) .. controls (322.38,65.61) and (314.42,72.82) .. (304.6,72.82) .. controls (294.78,72.82) and (286.82,65.61) .. (286.82,56.73) -- cycle ;

\draw (207.46,94.05) node [anchor=north west][inner sep=0.75pt]  [font=\Large]  {$\overline{l}$};
\draw (55.02,96.05) node [anchor=north west][inner sep=0.75pt]  [font=\Large]  {$e$};
\draw (53.37,6.74) node [anchor=north west][inner sep=0.75pt]  [font=\normalsize]  {$H^{( \dagger )}$};
\draw (121.45,99.15) node [anchor=north west][inner sep=0.75pt]    {$\text{\textcolor[rgb]{0.29,0.56,0.89}{cut}}\textcolor[rgb]{0.29,0.56,0.89}{_{24}}$};
\draw (122.33,87.97) node [anchor=north west][inner sep=0.75pt]    {$\text{\textcolor[rgb]{0.29,0.56,0.89}{cut}}\textcolor[rgb]{0.29,0.56,0.89}{_{23}}$};
\draw (121.45,110.32) node [anchor=north west][inner sep=0.75pt]    {$\text{\textcolor[rgb]{0.29,0.56,0.89}{cut}}\textcolor[rgb]{0.29,0.56,0.89}{_{25}}$};
\draw (212.23,7.53) node [anchor=north west][inner sep=0.75pt]  [font=\normalsize]  {$H^{( \dagger )}$};
\draw (217.52,39.47) node [anchor=north west][inner sep=0.75pt]  [font=\normalsize]  {$H^{( \dagger )}$};
\draw (433.39,57.18) node [anchor=north west][inner sep=0.75pt]  [font=\Large]  {$\overline{l}$};
\draw (429.21,97.32) node [anchor=north west][inner sep=0.75pt]  [font=\Large]  {$e$};
\draw (263.42,5.94) node [anchor=north west][inner sep=0.75pt]  [font=\normalsize]  {$H^{( \dagger )}$};
\draw (331.49,98.35) node [anchor=north west][inner sep=0.75pt]    {$\text{\textcolor[rgb]{0.29,0.56,0.89}{cut}}\textcolor[rgb]{0.29,0.56,0.89}{_{35}}$};
\draw (332.37,87.17) node [anchor=north west][inner sep=0.75pt]    {$\text{\textcolor[rgb]{0.29,0.56,0.89}{cut}}\textcolor[rgb]{0.29,0.56,0.89}{_{34}}$};
\draw (331.49,109.53) node [anchor=north west][inner sep=0.75pt]    {$\text{\textcolor[rgb]{0.29,0.56,0.89}{cut}}\textcolor[rgb]{0.29,0.56,0.89}{_{45}}$};
\draw (422.27,6.74) node [anchor=north west][inner sep=0.75pt]  [font=\normalsize]  {$H^{( \dagger )}$};
\draw (271.01,91.29) node [anchor=north west][inner sep=0.75pt]  [font=\normalsize]  {$H^{( \dagger )}$};
\draw (482.46,101.43) node [anchor=north west][inner sep=0.75pt]  [font=\Large]  {$\overline{l}$};
\draw (646.32,113.89) node [anchor=north west][inner sep=0.75pt]  [font=\Large]  {$e$};
\draw (474.34,17.39) node [anchor=north west][inner sep=0.75pt]  [font=\normalsize]  {$H^{\dagger }$};
\draw (543.3,98.63) node [anchor=north west][inner sep=0.75pt]    {$\text{\textcolor[rgb]{0.29,0.56,0.89}{cut}}\textcolor[rgb]{0.29,0.56,0.89}{_{14}}$};
\draw (633.2,18.19) node [anchor=north west][inner sep=0.75pt]  [font=\normalsize]  {$H$};
\draw (638.49,50.13) node [anchor=north west][inner sep=0.75pt]  [font=\normalsize]  {$H$};

\end{tikzpicture}

\caption{All the double cuts of the one-loop amplitude that contributes to $\dot{C}_{eH}$. The gray blobs are tree-level on-shell amplitudes.
}\label{fi:2-cut}
\end{figure}
\end{widetext}

\section{Example: $\gamma_{He,eH}$}\label{sec:CalcAnomal}
We demonstrate the computation of anomalous dimension matrix by an example in the SMEFT: the operator $\mathcal{O}_{He}=(\bar{e}\gamma^{\mu}e)(H^{\dagger}i\overleftrightarrow{D}_{\mu}H)$ contributing to the renormalization of $\mathcal{O}_{eH}=(\bar{l}eH)(H^{\dagger}H)$.
The relevant Lagrangian terms are
\begin{align}\begin{split}
    \mathcal{L}=&|D_{\mu}H|^2+i\,\bar{l}D\!\!\!\!/\, l+i\,\bar{e}D\!\!\!\!/\, e- Y_e (\bar{e}lH^{\dagger})- Y_e^{\dagger} (\bar{l}eH)\\
    &-\lambda(H^{\dagger}H)^2+C_{He}\mathcal{O}_{He}+C_{eH}\mathcal{O}_{eH}+\dots 
\end{split}\end{align}
The anomalous dimension $\gamma$ resides in the one-loop amplitude as the coefficient of the UV divergence
\eq{
\mathcal{A}^{\rm 1-loop}&\left(1_{\bar{l}^i},2_{e},3_{H_j},4_{H^{\dagger k}},5_{H_l}\right) = \frac{\gamma}{\epsilon}\mc{B}_{eH} + O(\epsilon^0)\,.
}
There is only one available form of local amplitude in the UV divergence, $\mc{B}_{eH} = \left( \delta^i_j\delta^k_l+\delta^i_l\delta^k_j\right)[12]\,,$
which corresponds to the operator $\mc{O}_{eH}$.

There are non-trivial double cuts of this one-loop amplitude in the following channels
\eq{
    \mc{I} \in\{ (23),(24),(25),(14),(34),(35),(45) \}\,,
}
as shown in the Fig.~\ref{fi:2-cut}.
The unique structure in $\mc{B}_{eH}$ makes the matrix $\mc{K}^{\rm (jf)}$ in eq.~\eqref{eq:adm_master} trivial: for each of the above channel, there is only one contributing $J$ and no degeneracy
\begin{itemize}
    \item $J=1/2$ for channels $(23)$, $(24)$, $(25)$, $(14)$.
    
    \item $J=0$ for channels $(34)$, $(35)$, $(45)$.
\end{itemize}
Therefore, the only missing pieces are the partial wave coefficients $\mc{M}_{\rm L,R}^J$ for each of the channels, with the particular $J$ indicated above. There may be non-zero partial waves for other $J$ in the expansion, which, however, must not contribute due to the selection rule of angular momentum conservation.
We show a sample of the result in table~\ref{tab:pwbasis}, while the full result is derived and presented in the appendix. We have put the $SU(2)$ gauge factor into the j-basis, which makes sure that $\mc{B}_{eH}$ is reproduced from eq.~\eqref{eq:PWBIntegral} including the gauge sector.
For example, the $\mc{A}_{\rm R}$ for the cut $\mc{I}=(24)$ is expanded as an infinite series
\eq{\small
    \mc{A}_{\rm R} 
    &= 2Y_e^\dagger\lambda\left(\delta^{j}_{i}\delta^l_{k'}+\delta^{j}_{k'}\delta^l_i\right)\frac{1}{\vev{1'1}} \\
    &=\mathcal{M}^{1/2}_{\rm R}\mathcal{B}^{1/2} 
    +\mathcal{M}^{3/2}_{\rm R}\mathcal{B}^{3/2} +\dots 
}
But we don't need to compute the $J\ge 3/2$ components because we know that $\mc{K}^{\rm (jf)}$ in eq.~\eqref{eq:adm_master} is non-zero only for $J=1/2$.
One may also notice that $\mc{M}^{1/2}_{\rm R} \sim \frac{2}{s_{14}+s_{15}}$ in this case. Such rational functions will be cancelled once all the unitarity cuts are summed up, which reflects the fact that these contributions are correlated as they come from the same Feynman diagram.

$\text{cut}_{14}$ is a much more non-trivial case, while the $J=1/2$ partial-wave basis of $\mc{A}_{\rm R}$ is not unique. 
\begin{align}\small\label{eq:cut14}
    \frac{\langle 3|5-2'|2]}{\vev{1'3}}=[1'2]-
    \mc{M}^{1/2}_{{\rm R}2}\mc{B}^{1/2}_2+\sum_{J>1/2}...
\end{align}
where $\mc{B}^{1/2}_2=\frac{1}{s_{235}}[1'3][25]\vev{35}$ has $J_{35}=1$, which is incompatible with $\mc{B}_{eH}$ with $J_{35}=0$. Therefore, only the first term in eq.~\eqref{eq:cut14} contributes to table~\ref{tab:pwbasis}. 

One might notice that $\text{cut}_{34}$ and $\text{cut}_{45}$ with the intermediate states $\{\bar{e},e\}$ should not have the $J=0$ component in their partial-wave expansion due to the bound $J\ge|\Delta h|=1$. In this case, both sides of the unitarity cut contain the IR divergence, but the integrations have nontrivial results with angular momentum $J_{34}=0$ or $J_{45}=0$. The corresponding partial-wave expansions in table~\ref{tab:pwbasis} are obtained after doing the BCFW shift. More details of the calculation are in the supplement. Surprisingly, these two unitarity cuts, exactly cancel with $\text{cut}_{25}$ and $\text{cut}_{23}$ in terms of contribution to the anomalous dimension of $\mc{B}_{eH}$. The deeper reason for the cancellation and IR divergence will be discussed in future work.

After plugging all CG coefficients in eq.~\eqref{eq:adm_master}, we obtain 
\begin{align}
    \gamma_{He,eH}=-\frac{\lambda}{4\pi^2}Y_e^{\dagger}-\frac{1}{8\pi^2}[Y_e^{\dagger}Y_eY_e^{\dagger}]+\dots
\end{align}
which agrees with \cite{Jenkins:2013zja,Jenkins:2013wua}.

\begin{table}[htbp]
    \centering
    \begin{align*}
    \begin{array}{|c|l|l|}
        \hline
        \multicolumn{3}{|c|}{\text{channel }\blue{(23)}\text{ with mediate state }\blue{\{e,H\}, J=1/2}}\\
        \hline
        \mc{B} & \mc{B}_{\rm L}=-s_{23}^{-1}\delta^j_{j'} [2|3|1'\rangle & \mc{B}_{\rm R}= [1'1]\left( \delta_i^{j'}\delta_k^l+\delta_i^l\delta_k^{j'}\right) \\
        \mc{M} & \mc{M}_{\rm L}=2C_{He}s_{23} & \mc{M}_{\rm R}=(2Y_e^{\dagger}\lambda +\frac12[Y^{\dagger}_eY_eY^{\dagger}_e])\frac{2}{s_{14}+s_{15}}  \\
        \hline\hline
        \multicolumn{3}{|c|}{\text{channel }\blue{(24)}\text{ with mediate state }\blue{\{e,H\}, J=1/2}}\\
        \hline
        \mc{B} & \mc{B}_{\rm L}=-s_{24}^{-1} [2|4|1'\rangle\delta^{k'}_{k} & \mc{B}_{\rm R}= [1'1]\left( \delta_i^{j}\delta_{k'}^l+\delta_i^l\delta_{k'}^{j}\right) \\
        \mc{M} & \mc{M}_{\rm L}=-2C_{He}s_{24} & \mc{M}_{\rm R}=2Y_e^{\dagger}\lambda  \frac{2}{s_{13}+s_{15}}  \\
        \hline\hline
        \multicolumn{3}{|c|}{\text{channel }\blue{(14)}\text{ with mediate state }\blue{\{e,H\}, J=1/2}}\\
        \hline
        \mc{B} & \mc{B}_{\rm L}=s_{14}^{-1} [1|4|1'\rangle\delta^{i'}_{i}\delta^{k'}_k & \mc{B}_{\rm R}= [1'2]\left( \delta_{i'}^{j}\delta_{k'}^{l}+\delta_{i'}^{l}\delta_{k'}^{j}\right) \\
        \mc{M} & \mc{M}_{\rm L}=-2[Y^{\dagger}_eY_e] & \mc{M}_{\rm R}=-C_{He}Y_e^{\dagger}  \\
        \hline\hline
        \multicolumn{3}{|c|}{\text{channel }\blue{(45)}\text{ with mediate state }\blue{\{\bar{e},e\}, J=0}}\\
        \hline
        \mc{B} & \mc{B}_{\rm L}=\delta^{l}_{k}\delta^{k'}_{l'}+\delta^{k'}_{k}\delta^{l}_{l'} & \mc{B}_{\rm R}=[12]\delta^{l'}_{i}\delta_{k'}^{j}  \\
        \mc{M} & \mc{M}_{\rm L}=[Y^{\dagger}_eY_e] & \mc{M}_{\rm R}=[Y_e^{\dagger}C_{He}]  \frac{s_{23}}{s_{12}+s_{13}}  \\
        \hline
    \end{array}
    \end{align*}
    \caption{Partial wave components relevant for the computation of $\gamma_{He,eH}$. We list part of the cut channels, intermediate states, and the angular momenta that are summed over in eq.~\eqref{eq:adm_master}. In each case, we present the partial wave bases $\mc{B}_{\rm L/R}$ and the coefficients $\mc{M}_{\rm L/R}$ of the left/right sub-amplitudes. The square brackets around the couplings denote flavor matrix multiplication. The full list is given in the supplementary material.}
    \label{tab:pwbasis}
\end{table}

\comment{
\begin{widetext}
\input{pw_tab_abridge}
\end{widetext}
}

\section{Conclusion}\label{sec:conclusion}

We constructed the general partial wave basis for scattering $N\to M$ massless or massive particles based on the Poincar\'e Clebsch-Gordan Coefficients that define the irreducible multi-particle states. We closely examined them by introducing the inner product, which induces orthonormality of the partial wave basis, a generalization for the normalized Wigner d-matrices in the particular case of $2\to 2$ scattering. We discovered a finite number of degenerate partial waves, up to functions of invariant kinematics, for a given angular momentum $J$ when the number of particles $N>2$. 
We studied the expansion of arbitrary IR finite amplitudes into partial waves, which provides an algebraic way of integrating on-shell phase space. The technique is applied to computations of anomalous dimensions for effective operators at one-loop level, where selection rules due to angular momentum conservation \cite{Jiang:2020rwz} are implied by the orthogonality of the Poincar\'e CGC, with possible IR divergence in the loop integrals removed by BCFW momenta shifts.

In a broader sense, partial wave expansion may benefit all kinds of on-shell phase space integration, including the computation of cross section and the unitarity cuts. The inner product of Poincar\'e CGC can be easily extended to $N>2$ -particle states, which are relevant for $N>2$ cuts for multi-loop diagrams. An algebraic method of such computations can open a new gate for the study of loop level amplitudes. 
Phenomenologically, instead of reading only the energy spectrum of a process in the traditional analysis, one could also extract the spin information from the different partial wave spectral functions $\mc{M}^J$. We also expect more unitarity and positivity bound for effective field theories when the partial wave information is included \cite{Arkani-Hamed:2020blm,Bern:2021ppb}. 

\section{Acknowledgements}
J.S. is supported by the National Natural Science Foundation of China
under Grants No. 12025507, No. 11690022, No.11947302; and is supported by the Strategic Priority Research Program and Key Research Program of Frontier Science of the Chinese Academy
of Sciences under Grants No. XDB21010200, No. XDB23010000, and No. ZDBS-LY-7003 and
CAS project for Young Scientists in Basic Research YSBR-006. M.-L.X. and Y.-H.Z. acknowledge helpful conversation with Jiang-Hao Yu. M.-L.X. is supported in part by the U.S. Department of Energy under contracts No. DE-AC02-06CH11357 at Argonne and No.DE-SC0010143 at Northwestern.

%

\end{document}